\documentclass[12pt,a4paper]{article}
\usepackage[T1]{fontenc}        
\usepackage[dvips]{epsfig}
\title{
  {\vspace{-2cm} \normalsize
     \epsfig{figure=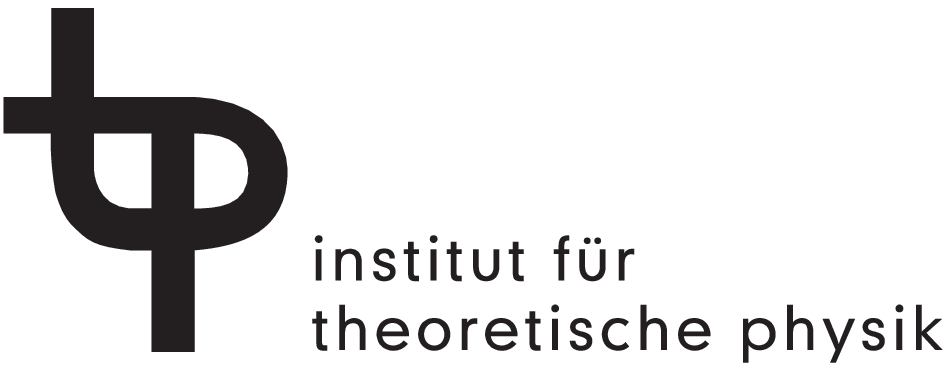,width=80mm}
     \hfill\parbox[b][30mm][t]{35mm}{MS-TP-04-1 \\
                                     DESY 04-015 \\
                                     hep-lat/0402003}  }\\[25mm]
  Chiral perturbation theory\\
  for partially quenched twisted mass lattice QCD
  }
\author{Gernot M\"unster, Christian Schmidt\\
        Institut f\"ur Theoretische Physik,
        Universit\"at M\"unster\\
        Wilhelm-Klemm-Str.~9, D-48149 M\"unster, Germany\\
        e-mail: munsteg@uni-muenster.de, sschmidt@uni-muenster.de\\[1em]
        Enno Scholz\\
        Deutsches Elektronen-Synchrotron DESY\\
        Notkestr.~85, D-22603 Hamburg, Germany\\
        e-mail: enno.e.scholz@desy.de}
\date{February 3, 2004\\(revised: October 01, 2004)}
\newcommand{\I}{\ensuremath{\mathrm{i}\,}}
\newcommand{\E}{\ensuremath{\mathrm{e}\,}}
\newcommand{\sTr}{\mbox{s\kern-.08em Tr}}
\newcommand{\rc}[1]{\rho\cos\omega_{#1}}
\begin{document}
\maketitle

\begin{abstract}
Partially quenched Quantum Chromodynamics with Wilson fermions on a
lattice is considered in the framework of chiral perturbation theory in
the mesonic sector. Two degenerate quark flavours are associated with a
chirally twisted mass term. The pion masses and decay constants are
calculated in next-to-leading order including terms linear in the
lattice spacing $a$.\\[5mm]
\end{abstract}
%

A notorious problem for numerical simulations of Quantum Chromodynamics
with dynamical quarks is to reach the region of realistic small quark
masses.  To deal with this problem different methods are being used. 
On the theoretical side, chiral perturbation theory supplies us with
formulae which allow to extrapolate pion masses and other observables
from medium to small quark masses \cite{Weinberg,GL1,GL2}.  The low
energy parameters of chiral perturbation theory, the
\textit{Gasser-Leutwyler coefficients}, can in turn be determined by
numerical simulations of lattice QCD, see \cite{Wittig} for a review. 
The effects of the finite lattice spacing $a$ are taken into account in
chiral perturbation theory in form of an expansion in powers of $a$
\cite{ShaSi,LeeSha,Rupak-Shoresh,Aoki,BRS}. In \cite{ALPHA,UKQCD,
Fleming,qqq1,qqq2,qqq3} numerical results from lattice QCD are compared
with chiral perturbation theory in next-to-leading order.

On the algorithmic side partially quenched QCD is an approach to the
regime of small quark masses.  In a numerical simulation of partially
quenched QCD the Monte Carlo updates are being made with sea-quarks,
which have large enough masses $m_S$ in order to allow a tolerable
simulation speed.  On the other hand, quark propagators and related
observables are evaluated with smaller valence-quark masses $m_V$.
Chiral perturbation theory has been adopted to the case of partially
quenched QCD in \cite{BG,SharpePQ}.

Recently it has been advocated to employ a chirally twisted quark mass
matrix for Wilson fermions \cite{TM,Frezzotti} in order to improve the
efficiency of QCD simulations \cite{FR1,FR2}. At present, this approach
appears to be very attractive for reducing lattice effects in numerical
simulations. First numerical results using this approach are very
promising \cite{Jansen}. The corresponding modifications of chiral
perturbation theory for lattice QCD have been worked out to order $a$
in \cite{MS}.

Combining these approaches, which is a logical next step, it appears
attractive to simulate QCD with chirally twisted quark masses in a
partially quenched manner.  For the theoretical analysis of the data it
is desirable to extend the results of \cite{MS} to the partially
quenched case.  This is the object of this letter.

We consider lattice QCD with $N_f =2$ flavours of sea quarks and the
same number of valence quarks.  For simplicity we restrict ourselves to
the case of degenerate quark masses ($m_u=m_d=m$).  The quark mass
matrix contains two chiral twist angles.  Chiral perturbation theory is
applied to the mesonic sector. The pion masses and decay constants are
calculated in chiral perturbation theory in next-to-leading (one-loop)
order, including lattice terms linear in the lattice spacing $a$.

Recent Monte Carlo simulations of the partially quenched model
\cite{qqq2,qqq3} show that the one-loop contributions of chiral
perturbation theory (for vanishing twist) can be identified in the
numerical data and used to estimate Gasser-Leutwyler coefficients.
In particular, it turned out that partially quenched chiral
perturbation theory for lattice QCD is already applicable for
quark masses below $\frac{1}{2} m_s$.

A theoretical description of partially quenched QCD can be obtained
through the introduction of ghost quarks \cite{Morel}. For each valence
quark a corresponding bosonic ghost quark is added to the model. The
functional integral over the ghost quark fields then cancels the
fermion determinant of the valence quarks and only the sea quark
determinant remains in the measure. In our case there are 2 flavours of
valence quarks, sea quarks and ghost quarks, each. The quark mass
matrix is
\begin{equation}
M = \mbox{diag}(m_V, m_V, m_S , m_S, m_V, m_V).
\end{equation}
A chirally twisted mass matrix, depending on two twist angles can be
introduced in the form
\begin{equation}
M(\omega_V,\omega_S) = M \E^{\I \omega_V \tau_3^V \gamma_5}\
\E^{\I \omega_S \tau_3^S \gamma_5}\
\E^{\I \omega_V \tau_3^G \gamma_5}\,,
\end{equation}
where
\begin{equation}
\tau_3^V=
\left( \begin{array}{ccc}
\tau_3 & 0 & 0 \\
0 & 0 & 0 \\
0 & 0 & 0
\end{array} \right),
\qquad
\tau_3^S=
\left( \begin{array}{ccc}
0 & 0 & 0 \\
0 & \tau_3 & 0 \\
0 & 0 & 0
\end{array} \right),
\qquad
\tau_3^G=
\left( \begin{array}{ccc}
0 & 0 & 0 \\
0 & 0 & 0 \\
0 & 0 & \tau_3
\end{array} \right).
\end{equation}

In chiral perturbation theory the dynamics of the pseudo-Goldstone
fields is described by a low energy effective Lagrangian. For partially
quenched QCD the pseudo-Goldstone fields, which we shall generally call
pions, are parameterized by a graded matrix of the form
\begin{equation}
U = \left( \begin{array}{cc}
A & B \\
C & D
\end{array} \right).
\end{equation}
The $4 \times 4$ matrix $A$ and the $2 \times 2$ matrix $D$ contain
commuting numbers, whereas the $4 \times 2$ matrix $B$ and the $2
\times 4$ matrix $C$ have anticommuting entries. $U$ is an element of
the supergroup SU(4|2) and can be represented as
\begin{equation}
U(x) = \exp \left(\frac{\I}{F_0} \Phi(x) \right).
\end{equation}
The commuting elements of the graded matrix $\Phi$ represent the
pseudo-Goldstone bosons made from a quark and an anti-quark with equal
statistics, and the anticommuting elements of $\Phi$ represent
pseudo-Goldstone fermions which are built from one fermionic quark and
one bosonic quark. The supertrace of $\Phi$ has to vanish,
\begin{equation}
\sTr\, \Phi = 0\,,
\end{equation}
where one defines
\begin{equation}
\sTr
\left( \begin{array}{cc}
A & B \\
C & D
\end{array} \right)
= \mbox{Tr} A - \mbox{Tr} D\,.
\end{equation}
This condition introduces technical complications in loop calculations.
One way to deal with them \cite{Sharpe-Shoresh} is to drop the
condition $\sTr\, \Phi = 0$, adding a mass term proportional to
$(\sTr\, \Phi)^2$ and sending its mass to infinity in the end.
Another way, which we followed in our calculations, is to choose a
basis of 35 super-traceless generators $T_j$ of SU(4|2) and to expand
\begin{equation}
\Phi = \sum_j \pi_j T_j \,.
\end{equation}

The chiral effective Lagrangian to leading order, including lattice
artifacts up to first order in the lattice spacing $a$ is given by
\cite{Rupak-Shoresh}
\begin{equation}
\mathcal{L}_2
= \frac{F_0^2}{4}\,\sTr \left( \partial_\mu U^\dagger \,
\partial^\mu U \right)
- \frac{F_0^2}{4}\,\sTr \left( \chi U^\dagger + U \chi^\dagger
\right)
- \frac{F_0^2}{4}\,\sTr \left( \rho U^\dagger + U \rho^\dagger
\right).
\end{equation}
The mass term contains the matrix
\begin{equation}
\chi = 2 B_0 M
\end{equation}
and the lattice term is parameterized by
\begin{equation}
\rho = 2 W_0 a \, {\bf 1}\,,
\end{equation}
with two constants $B_0$ and $W_0$.

Similar to the unquenched case, the chiral twist can be incorporated in
chiral perturbation theory by transforming the lattice term into
\begin{equation}
\rho(\omega) \equiv \rho(\omega_V, \omega_S) =
\rho \, \E^{\I \omega_V \tau_3^V}\,
\E^{\I \omega_S \tau_3^S}\,\E^{\I \omega_V \tau_3^G}\,.
\end{equation}
The effective Lagrangian in next-to-leading order (with respect to
masses and momenta in the usual Weinberg power counting scheme) is then
given by
\cite{Rupak-Shoresh}
\begin{eqnarray}
\mathcal{L}_4 & = & \frac{F_0^2}{4}\,\sTr \left( \partial_\mu U
\partial_\mu U^\dagger \right)
- \frac{F_0^2}{4}\,\sTr \left( \chi U^\dagger + U \chi^\dagger \right)
- \frac{F_0^2}{4}\,\sTr \left( \rho(\omega) U^\dagger
+ U \rho(\omega)^\dagger \right) \\
& & - L_1 \left[ \sTr \left( \partial_\mu U \partial_\mu U^\dagger
\right) \right]^2
- L_2\, \sTr \left( \partial_\mu U \partial_\nu U^\dagger \right)
\sTr \left( \partial_\mu U \partial_\nu U^\dagger \right) \nonumber \\
& & - L_3\, \sTr \left( \left[ \partial_\mu U \partial_\mu  U^\dagger
\right]^2 \right)
+ L_4\, \sTr \left( \partial_\mu U \partial_\mu  U^\dagger \right)
\sTr \left( \chi U^\dagger + U \chi^\dagger \right) \nonumber \\
& & +  W_4\, \sTr \left( \partial_\mu U \partial_\mu  U^\dagger \right)
\sTr \left( \rho(\omega) U^\dagger + U \rho(\omega)^\dagger \right)
+ L_5\, \sTr \left( \partial_\mu U \partial_\mu  U^\dagger
\left[ \chi U^\dagger + U \chi^\dagger \right] \right) \nonumber \\
& & + W_5\, \sTr \left( \partial_\mu U \partial_\mu  U^\dagger
\left[ \rho(\omega) U^\dagger + U \rho(\omega)^\dagger \right] \right)
- L_6 \left[ \sTr \left( \chi U^\dagger + U \chi^\dagger \right)
\right]^2 \nonumber \\
& & - W_6\, \sTr \left( \chi U^\dagger + U \chi^\dagger \right)
\sTr \left( \rho(\omega) U^\dagger + U \rho(\omega)^\dagger \right)
- L_7 \left[ \sTr \left( \chi U^\dagger - U \chi^\dagger \right)
\right]^2 \nonumber \\
& & - W_7\, \sTr \left( \chi U^\dagger - U \chi^\dagger \right)
\sTr \left( \rho(\omega) U^\dagger - U \rho(\omega)^\dagger \right)
- L_8\, \sTr \left( \chi U^\dagger \chi U^\dagger
+ U \chi^\dagger U \chi^\dagger \right) \nonumber \\
& & - W_8\, \sTr \left( \chi U^\dagger \rho(\omega) U^\dagger
+ U \rho(\omega)^\dagger U \chi^\dagger \right)
+ \mathcal{O} (a^2). \nonumber
\end{eqnarray}
In order to calculate the pion masses and other physical quantities,
the Lagrangian has to be expanded in terms of the fields contained in
$\Phi$. In the untwisted case the expansion is around the minimum (more
precisely, the saddle point) at vanishing $\Phi$. The chiral twist,
however, introduces a shift of the expansion point, as was already
observed in the unquenched case \cite{MS}. We have calculated this
shift in next-to-leading order including terms linear in the lattice
spacing $a$, and expanded the effective action around the shifted
minimum.

From the resulting expression the propagators and pion decay constants
can be calculated in a one-loop calculation. The results are known for
the case of vanishing twist \cite{Rupak-Shoresh}. Therefore one has to
account for the effects of the twist angles now. The lattice term
$\sTr ( \rho(\omega) U^\dagger + U \rho(\omega)^\dagger )$
yields the quadratic contribution
$\sTr (( \rho(\omega) + \rho(\omega)^\dagger ) \Phi^2)$, which
contains the combination
\begin{equation}
\rho(\omega) + \rho(\omega)^\dagger = 4 W_0 a (
\cos \omega_V {\bf 1}^V + \cos \omega_S {\bf 1}^S
+ \cos \omega_V {\bf 1}^G)\,.
\end{equation}
We see that in this term the twist amounts to a multiplication of the
untwisted lattice terms by factors $\cos \omega_V$ and $\cos \omega_S$,
respectively. We have checked all terms proportional to the
coefficients $W_i$ and found that for them the same rule holds.

In the one-loop calculation the propagators and vertices from the
leading order effective Lagrangian enter. We considered these terms and
found that for the propagator and 4-point vertices the twist produces
$\cos \omega$-factors as above. The 3-point vertices are modified by
$\sin \omega$-factors. The relevant Feynman diagrams, however, contain
two of these vertices and thus contribute to order $a^2$ only.

To summarize, at order $a$ the chiral twist amounts to adding factors
$\cos \omega_V$ or $\cos \omega_S$ to the untwisted quantities. This
rule will not apply to order $a^2$, where new vertices appear. Results
for masses and decay constants for partially quenched QCD with $N_f
= 3$ flavours have been given by Rupak and Shoresh \cite{Rupak-Shoresh}.
The formulae for a general number $N_S$ of sea quarks and $N_V = 2$
valence quarks are presented in \cite{qqq1}. They give masses and decay
constants for flavour-charged pions. From these we obtain the results
for the twisted case under consideration using the procedure explained
above. For the the pion masses we find
\begin{eqnarray}
m^2_{SS}&=&\chi_S + \rc{S}\nonumber\\
&&+ \frac{1}{32\pi^2 F_0^2}(\chi_S + \rc{S})^2
\log\left[\frac{\chi_S + \rc{S}}{\Lambda^2}\right] \nonumber\\
&&+ \frac{16}{F_0^2} (2L_6 - L_4)\chi_S^2
+ \frac{8}{F_0^2} (2L_8 - L_5)\chi_S^2
+ \frac{16}{F_0^2} (W_6 - L_4)\chi_S \rc{S}\nonumber\\
&&+ \frac{16}{F_0^2} (W_6 - W_4)\chi_S \rc{S}
+ \frac{8}{F_0^2} (2W_8 - W_5 - L_5)\chi_S \rc{S}
\end{eqnarray}
\begin{eqnarray}
m^2_{VV}&=&\chi_V + \rc{V}\nonumber\\
&&+\frac{1}{32\pi^2 F_0^2}(\chi_V + \rc{V})
\Bigg\{\chi_V - \chi_S + \rc{V} - \rc{S} \nonumber\\
&&\:+ \Big(2\chi_V - \chi_S + 2\rc{V} - \rc{S} \Big)
\log\left[\frac{\chi_V + \rc{V}}{\Lambda^2}\right]\Bigg\}\nonumber\\
&&+ \frac{16}{F_0^2} (2L_6 - L_4)\chi_S \chi_V
+ \frac{8}{F_0^2} (2L_8 - L_5)\chi_V^2
+ \frac{16}{F_0^2} (W_6 - L_4)\chi_S \rc{V}\nonumber\\
&&+ \frac{16}{F_0^2} (W_6 - W_4)\chi_V \rc{S}
+ \frac{8}{F_0^2} (2W_8 - W_5 - L_5)\chi_V \rc{V}
\end{eqnarray}
\begin{eqnarray}
m^2_{VS}&=&\frac{1}{2}
(\chi_V + \chi_S + \rc{V} + \rc{S})\nonumber\\
&&+ \frac{1}{64\pi^2 F_0^2}(\chi_V + \chi_S + \rc{V} + \rc{S})
(\chi_V + \rc{V})\log\left[\frac{\chi_V + \rc{V}}{\Lambda^2}\right]
\nonumber\\
&&+ \frac{8}{F_0^2} (2L_6 - L_4)\chi_S (\chi_V + \chi_S)
+ \frac{2}{F_0^2} (2L_8 - L_5)(\chi_V + \chi_S)^2\nonumber\\
&&+ \frac{8}{F_0^2} (W_6 - L_4)\chi_S (\rc{V} + \rc{S})
+ \frac{8}{F_0^2} (W_6-W_4)(\chi_V + \chi_S)\rc{S}\nonumber\\
&&+ \frac{2}{F_0^2} (2W_8 - W_5 - L_5)(\chi_V +\chi_S)(\rc{V} + \rc{S})
\end{eqnarray}
and for the decay constants
\begin{eqnarray}
\frac{F_{SS}}{F_0}&=&1\:-\:\frac{1}{16\pi^2 F_0^2}
(\chi_S + \rc{S})\log\left[\frac{\chi_S + \rc{S}}{\Lambda^2}\right]
\nonumber\\
&&+ \frac{4}{F_0^2} L_5\chi_S + \frac{8}{F_0^2} L_4\chi_{S}
+ \frac{4}{F_0^2} W_5\rc{S} + \frac{8}{F_0^2} W_4\rc{S}
\end{eqnarray}
\begin{eqnarray}
\frac{F_{VV}}{F_0}&=&1\:-\:\frac{1}{32\pi^2 F_0^2}
\Big(\chi_V + \chi_S + \rc{V} + \rc{S}\Big)
\log\left[\frac{\chi_V + \chi_S + \rc{V} + \rc{S}}{2\Lambda^2}\right]
\nonumber\\
&&+ \frac{4}{F_0^2} L_5\chi_V + \frac{8}{F_0^2} L_4\chi_S
+ \frac{4}{F_0^2} W_5\rc{V} + \frac{8}{F_0^2} W_4\rc{S}
\end{eqnarray}
\begin{eqnarray}
\frac{F_{VS}}{F_0}&=&1\:-\:\frac1{64\pi^2 F_0^2}
\bigg\{2\Big(\chi_S + \rc{S}\Big)
\log\left[\frac{\chi_S + \rc{S}}{\Lambda^2}\right]\nonumber\\
&&\:+ \Big(\chi_V + \chi_S + \rc{V} + \rc{S}\Big)
\log\left[\frac{\chi_V + \chi_S + \rc{V} + \rc{S}}{2\Lambda^2}\right]
\bigg\}\nonumber\\
&&+ \frac{1}{128\pi^2 F_0^2}\Bigg\{\chi_V - \chi_S + \rc{V} - \rc{S}
- \Big(\chi_S + \rc{S}\Big)
\log\left[\frac{\chi_V + \rc{V}}{\chi_S + \rc{S}}\right]\Bigg\}
\nonumber\\
&&+ \frac{2}{F_0^2} L_5(\chi_V + \chi_S) + \frac{8}{F_0^2} L_4\chi_S
+ \frac{2}{F_0^2} W_5(\rc{V} + \rc{S}) + \frac{8}{F_0^2} W_4 \rc{S}\,,
\end{eqnarray}
where
\begin{equation}
\chi_V = 2 B_0 m_V\,, \qquad \chi_S = 2 B_0 m_S\,, \qquad
\rho = 2 W_0 a \,,
\end{equation}
and $\Lambda$ is the renormalization scale. It should be noted that the
axial vector current used for the definition of the pion decay constant
is the Noether current corresponding to an axial rotation, which
differs from other commonly used axial vector currents
by $\mathcal{O}(a)$ terms.

For maximal twist, $\omega_V = \omega_S = \pi/2$, the lattice artifacts
vanish to order $a$, as has been shown for lattice QCD in general by
Frezzotti and Rossi \cite{FR1,FR2} and checked in chiral perturbation
in \cite{MS}. Terms of order $a^2$ could be incorporated along the
lines of \cite{BRS}. Such a calculation, which is technically rather
involved, is in progress.

The expressions for the pion masses and decay constants in partially
quenched lattice QCD with a chiral twist can be used in the analysis of
numerical results from Monte Carlo calculations and will aid the
extrapolation to small quark masses.

We thank I.~Montvay, F.~Farchioni and R.~Frezzotti for discussions.

%
\end{document}